\documentclass[twocolumn,showpacs,aps,prl,superscriptaddress]{revtex4}

\usepackage{graphicx}
\usepackage{dcolumn}
\usepackage{amsmath}
\usepackage{epsfig}

\RequirePackage{xspace}

\usepackage{relsize}
\def\babar{\mbox{\slshape B\kern-0.1em{\smaller A}\kern-0.1em
    B\kern-0.1em{\smaller A\kern-0.2em R}}}

\def\g     {\ensuremath{\gamma}\xspace}

\def\piz   {\ensuremath{\pi^0}\xspace}

\def\pip   {\ensuremath{\pi^+}\xspace}
\def\pim   {\ensuremath{\pi^-}\xspace}

\def\Kbar  {\kern 0.2em\overline{\kern -0.2em K}{}\xspace}

\def\Kz    {\ensuremath{K^0}\xspace}
\def\Kzb   {\ensuremath{\Kbar^0}\xspace}
\def\KzKzb {\ensuremath{\Kz \kern -0.16em \Kzb}\xspace}
\def\Kp    {\ensuremath{K^+}\xspace}
\def\Km    {\ensuremath{K^-}\xspace}

\def\KpKm  {\ensuremath{\Kp \kern -0.16em \Km}\xspace}

\def\Kstarzb {\ensuremath{\Kbar^{*0}}\xspace}

\def\Dbar    {\kern 0.2em\overline{\kern -0.2em D}{}\xspace}
\def\Db      {\ensuremath{\Dbar}\xspace}
\def\Dz      {\ensuremath{D^0}\xspace}
\def\Dzb     {\ensuremath{\Dbar^0}\xspace}
\def\DzDzb   {\ensuremath{\Dz {\kern -0.16em \Dzb}}\xspace}
\def\Dp      {\ensuremath{D^+}\xspace}
\def\Dm      {\ensuremath{D^-}\xspace}

\def\DpDm    {\ensuremath{\Dp {\kern -0.16em \Dm}}\xspace}
\def\Dstar   {\ensuremath{D^*}\xspace}
\def\Dstarb  {\ensuremath{\Dbar^*}\xspace}
\def\Dstarz  {\ensuremath{D^{*0}}\xspace}

\def\Dstarp  {\ensuremath{D^{*+}}\xspace}

\def\Ds      {\ensuremath{D^+_s}\xspace}

\def\Dss     {\ensuremath{D^{*+}_s}\xspace}

\def\Bbar    {\kern 0.18em\overline{\kern -0.18em B}{}\xspace}

\def\BB      {\ensuremath{B\Bbar}\xspace} 
\def\Bz      {\ensuremath{B^0}\xspace}
\def\Bzb     {\ensuremath{\Bbar^0}\xspace}
\def\BzBzb   {\ensuremath{\Bz {\kern -0.16em \Bzb}}\xspace}
\def\Bu      {\ensuremath{B^+}\xspace}
\def\Bub     {\ensuremath{B^-}\xspace}
\def\Bp      {\ensuremath{\Bu}\xspace}
\def\Bm      {\ensuremath{\Bub}\xspace}

\def\BpBm    {\ensuremath{\Bu {\kern -0.16em \Bub}}\xspace}

\def\BorBbar    {\kern 0.18em\optbar{\kern -0.18em B}{}\xspace}
\def\DorDbar    {\kern 0.18em\optbar{\kern -0.18em D}{}\xspace}
\def\KorKbar    {\kern 0.18em\optbar{\kern -0.18em K}{}\xspace}

\mathchardef\Upsilon="7107
\def\Y#1S{\ensuremath{\Upsilon{(#1S)}}\xspace}

\def\FourS {\Y4S}

\mathchardef\Deltares="7101
\mathchardef\Xi="7104
\mathchardef\Lambda="7103
\mathchardef\Sigma="7106
\mathchardef\Omega="710A

\def\Deltabar{\kern 0.25em\overline{\kern -0.25em \Deltares}{}\xspace}
\def\Lbar{\kern 0.2em\overline{\kern -0.2em\Lambda\kern 0.05em}\kern-0.05em{}\xspace}
\def\Sigbar{\kern 0.2em\overline{\kern -0.2em \Sigma}{}\xspace}
\def\Xibar{\kern 0.2em\overline{\kern -0.2em \Xi}{}\xspace}
\def\Obar{\kern 0.2em\overline{\kern -0.2em \Omega}{}\xspace}
\def\Nbar{\kern 0.2em\overline{\kern -0.2em N}{}\xspace}
\def\Xb{\kern 0.2em\overline{\kern -0.2em X}{}\xspace}

\def\mes        {\mbox{$m_{\rm ES}$}\xspace}

\def\DeltaE     {\mbox{$\Delta E$}\xspace}

\newcommand{\tev}{\ensuremath{\mathrm{\,Te\kern -0.1em V}}\xspace}
\newcommand{\gev}{\ensuremath{\mathrm{\,Ge\kern -0.1em V}}\xspace}
\newcommand{\mev}{\ensuremath{\mathrm{\,Me\kern -0.1em V}}\xspace}
\newcommand{\kev}{\ensuremath{\mathrm{\,ke\kern -0.1em V}}\xspace}
\newcommand{\ev}{\ensuremath{\mathrm{\,e\kern -0.1em V}}\xspace}
\newcommand{\gevc}{\ensuremath{{\mathrm{\,Ge\kern -0.1em V\!/}c}}\xspace}
\newcommand{\mevc}{\ensuremath{{\mathrm{\,Me\kern -0.1em V\!/}c}}\xspace}
\newcommand{\gevcc}{\ensuremath{{\mathrm{\,Ge\kern -0.1em V\!/}c^2}}\xspace}
\newcommand{\mevcc}{\ensuremath{{\mathrm{\,Me\kern -0.1em V\!/}c^2}}\xspace}

\def\invfb   {\ensuremath{\mbox{\,fb}^{-1}}\xspace}

\def\mus  {\ensuremath{\rm \,\mus}\xspace}

\def\mus        {\ensuremath{\,\mu{\rm s}}\xspace}    

\def\to                 {\ensuremath{\rightarrow}\xspace}

\def\pep2{PEP-II}

\def\gsim{{~\raise.15em\hbox{$>$}\kern-.85em
          \lower.35em\hbox{$\sim$}~}\xspace}
\def\lsim{{~\raise.15em\hbox{$<$}\kern-.85em
          \lower.35em\hbox{$\sim$}~}\xspace}

\xspace

\newcommand{\jprlBase}       {Phys.\ Rev.\ Lett.\xspace}
\newcommand{\jprBase}        {Phys.\ Rev.\xspace}
\newcommand{\jplBase}        {Phys.\ Lett.\xspace}
\newcommand{\nimBaseA}       {Nucl.\ Instr.\ Meth.\xspace}

\newcommand{\nima}      [1]  {\nimBaseA~A~{\bf #1}}

\newcommand{\plb}       [1]  {\jplBase\ B~{\bf #1}}

\newcommand{\jprl}      [1]  {\jprlBase\ {\bf #1}}
\newcommand{\jprd}      [1]  {\jprBase\ D~{\bf #1}}

\def\jetset74   {\mbox{\tt Jetset \hspace{-0.5em}7.\hspace{-0.2em}4}\xspace}

\newcommand{\BABARPubYear}    {04}
\newcommand{\BABARPubNumber}  {24}

\newcommand{\SLACPubNumber} {10627}

\def\dssz{\ensuremath{D_{sJ}^{*}(2317)^+}}
\def\dsso{\ensuremath{D_{sJ}(2460)^+}}
\def\dsjp{\ensuremath{D^{(*)+}_{sJ}}}

\def\figurebox#1#2#3{%
    \def\arg{#3}%
    \ifx\arg\empty
    {\hfill\vbox{\hsize#2\hrule\hbox to #2{\vrule\hfill\vbox to #1{\hsize#2\vfill}\vrule}\hrule}\hfill}%
    \else
    {\hfill\epsfbox{#3}\hfill}%
    \fi}

\begin{document}

\preprint{\babar-PUB-\BABARPubYear/\BABARPubNumber}
\preprint{SLAC-PUB-\SLACPubNumber}

\begin{flushleft}
\babar-PUB-\BABARPubYear/\BABARPubNumber\\
SLAC-PUB-\SLACPubNumber\\
\end{flushleft}

\title{
{\large \bf Study of \boldmath{${B\rightarrow \dsjp \Db^{(*)}}$}
Decays }
}

%
\author{B.~Aubert}
\author{R.~Barate}
\author{D.~Boutigny}
\author{F.~Couderc}
\author{J.-M.~Gaillard}
\author{A.~Hicheur}
\author{Y.~Karyotakis}
\author{J.~P.~Lees}
\author{V.~Tisserand}
\author{A.~Zghiche}
\affiliation{Laboratoire de Physique des Particules, F-74941 Annecy-le-Vieux, France }
% ...
\collaboration{The \babar\ Collaboration}
\noaffiliation

\date{\today}

\begin{abstract}
We report a study of {$\dssz$} and {$\dsso$} meson production in
$B$ decays. We observe the decays {$B^+\rightarrow \dsjp
\overline{D}^{(*)0}$} and {$B^0\rightarrow \dsjp D^{(*)-}$} with
the subsequent decays {$\dssz\to \Ds \piz$}, {$\dsso \to \Ds
\gamma$}, and {$\dsso\to D_s^{*+} \piz$}. Based on a data sample of
122.1 million $\BB$ pairs collected with 
the \babar\ detector at the PEP-II $B$ factory, we obtain branching fractions
for these modes, including the previously unseen decays
{$B\rightarrow \dsjp D^{*}$}. In addition, we perform an angular
analysis of {$\dsso\to D^+_s\gamma$} decays to test the different
{$\dsso$} spin hypotheses.
\end{abstract}


\maketitle
The unexpected observation of a narrow {$D_s^+\pi^0$} resonance with a
mass of 2317\mevcc was recently reported by the \babar\ 
collaboration~\cite{Aubert:2003fg} and confirmed by the CLEO experiment~\cite{Besson:2003cp}.
CLEO observed a second {$D_s^{*+}\pi^0$} resonance
with a mass close to 2460\mevcc~\cite{Besson:2003cp},
previously suggested~\cite{Aubert:2003fg} 
and later confirmed~\cite{Aubert:2003pe} by \babar. 
The Belle collaboration confirmed
both resonances and found two additional decay modes  
for the higher-mass state~\cite{Abe:2003jk}, {$D_s^+\gamma$} and
{$D_s^+\pi^+\pi^-$}. These resonances are usually interpreted as
$P$-wave $c\bar{s}$ quark states~\cite{Cahn:2003cw,
Bardeen:2003kt, Godfrey:2003kg, Dai:2003yg, Deandrea:2003gb}, although other 
interpretations~\cite{Barnes:2003dj, vanBeveren:2003kd,
Cheng:2003kg, Szczepaniak:2003vy,  Browder:2003fk} cannot be ruled out, 
and will be referred to in the following as {$\dssz$} and {$\dsso$} 
mesons. 

The new states were first observed in $e^+e^-\to c
\bar c$ collisions. Their observation in exclusive $B\to
\dsjp\overline{D}^{(*)}$ decays allows additional properties of
the {$\dsjp$} states to be studied: the $\dsso\rightarrow
D^+_s\gamma$ helicity angle distribution in $B$ decays can be used
to obtain information on the $\dsso$ spin
$J$~\cite{Krokovny:2003zq}, and the measurement of the different
branching fractions can help clarify the nature of these states.

In this Letter we consider the $\dsjp$ production modes $\Bu\to
\dsjp \Dbar^{(*)0}$  and  $\Bz\to \dsjp D^{(*)-}$ with the
subsequent decays $\dssz\rightarrow D^+_s\piz$, 
$\dsso\rightarrow D^{*+}_s\piz$, and $\dsso\rightarrow D^+_s\gamma$. 
Our intention is to observe previously unseen decay chains, measure 
branching fractions for all channels, and determine the $\dsso$
spin by means of an angular analysis. Charge-conjugate reactions
are assumed throughout this paper.

The measurements reported here use 113\invfb\ of data,
corresponding to $(122.1 \pm 1.3)\times 10^6$ \BB pairs, collected
at the \FourS resonance  with the \babar\
detector~\cite{ref:detector} at the \pep2\ asymmetric-energy $B$
factory.

We reconstruct \Db and \Ds mesons in the modes $\Dzb\to\Kp\pim$,
 $\Kp\pim\piz$, $\Kp\pim\pip\pim$; $\Dm\to\Kp\pim\pim$; and
$\Ds\to\phi\pip$ ($\phi\to\Kp\Km$), $\Kstarzb\Kp$ ($\Kstarzb\to\Km
\pip$).
The reconstructed mass of the
\Db and \Ds candidates is required to be within $2.5\sigma$
($3\sigma$ for $\Kp\pim$, $\Kp\pim\pim$, and $\phi\pip$) of the
nominal $D$ masses, where the $D$ mass resolution $\sigma$, found 
in the data, is close to $12\mevcc$ for $\Db \to \Kp\pim\piz$ 
decays and varies from 5.3 to 6.3\mevcc for the other decay modes.

 The $D^*$ candidates are reconstructed in the decay modes $D^{*+}\to \Dz\pip$,
 $\Dstarz\to \Dz\piz$, $\Dz\g$, and $\Dss\to\Ds\gamma$. The mass
 difference $\Delta m$ between the \Dstar and $D$ candidates is required to
be within $2 \mevcc$ of its nominal value~\cite{ref:pdg2004} for
$\Dstarp \to \Dz\pip$ and $\Dstarz\to \Dz\piz$ ($10 \mevcc$ for
$\Dstarz \to \Dz\gamma$ and $\Dss\to\Ds\gamma$), corresponding to
about  $5\sigma_{\Delta m}$ for \Dstarp and $2\sigma_{\Delta m}$
for \Dstarz and \Dss.

\begin{figure*}[htb]
\begin{center}
 \hspace*{-0mm}\includegraphics[width=6.2 cm,height=8cm]{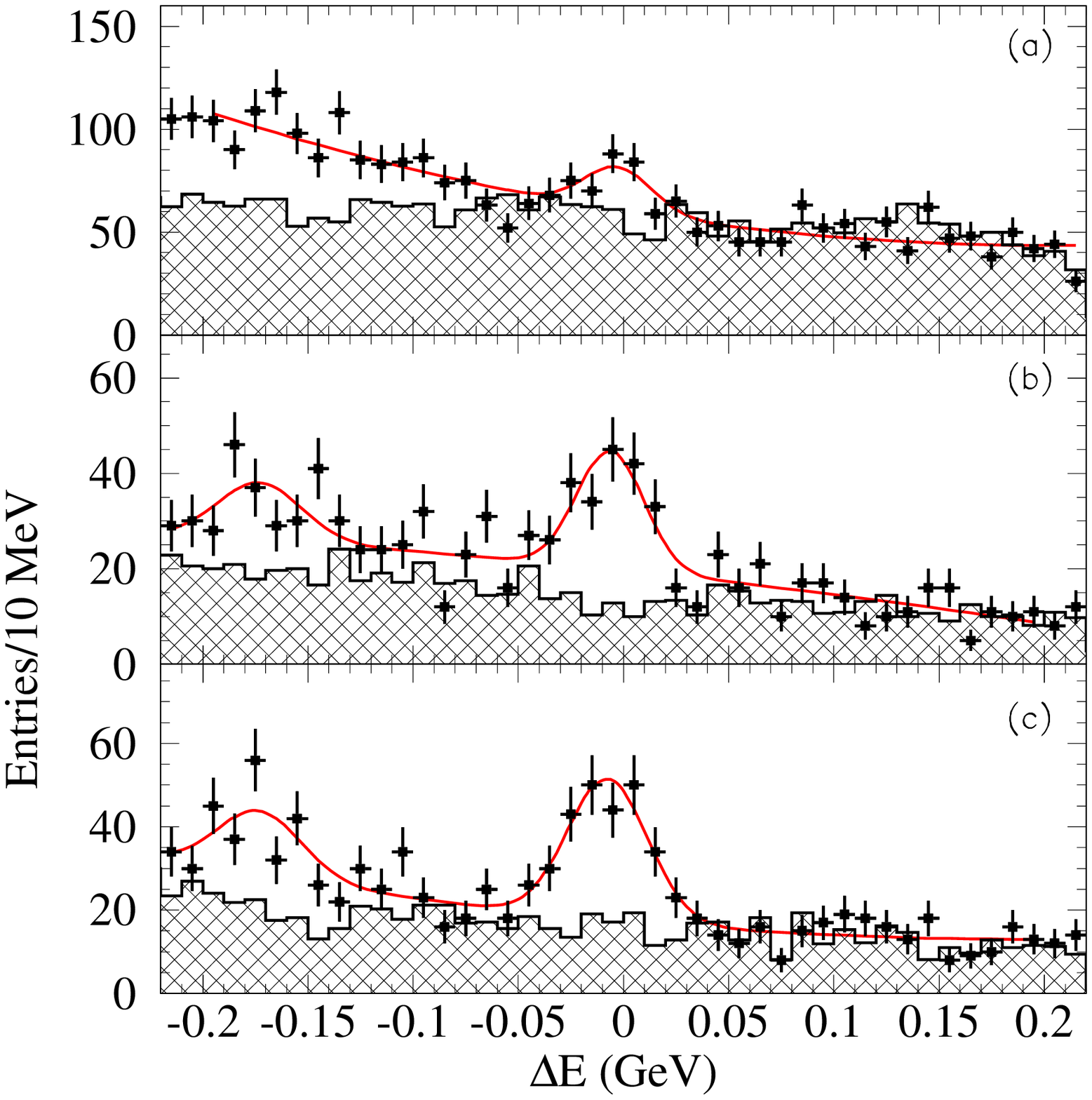}
 \hspace*{-5mm}\includegraphics[width=6.2 cm,height=8cm]{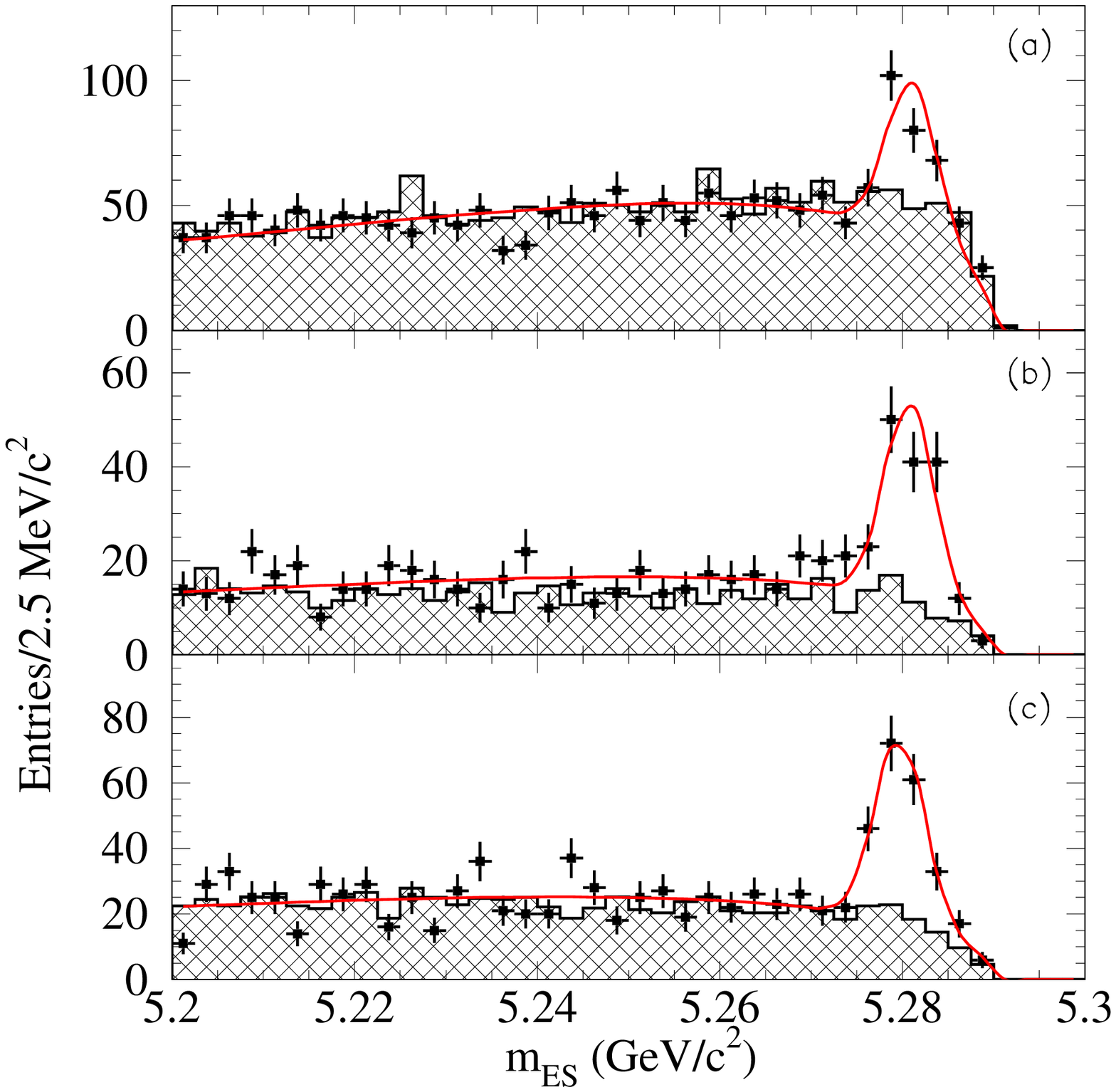}
 \hspace*{-5mm}\includegraphics[width=6.2 cm,height=8cm]{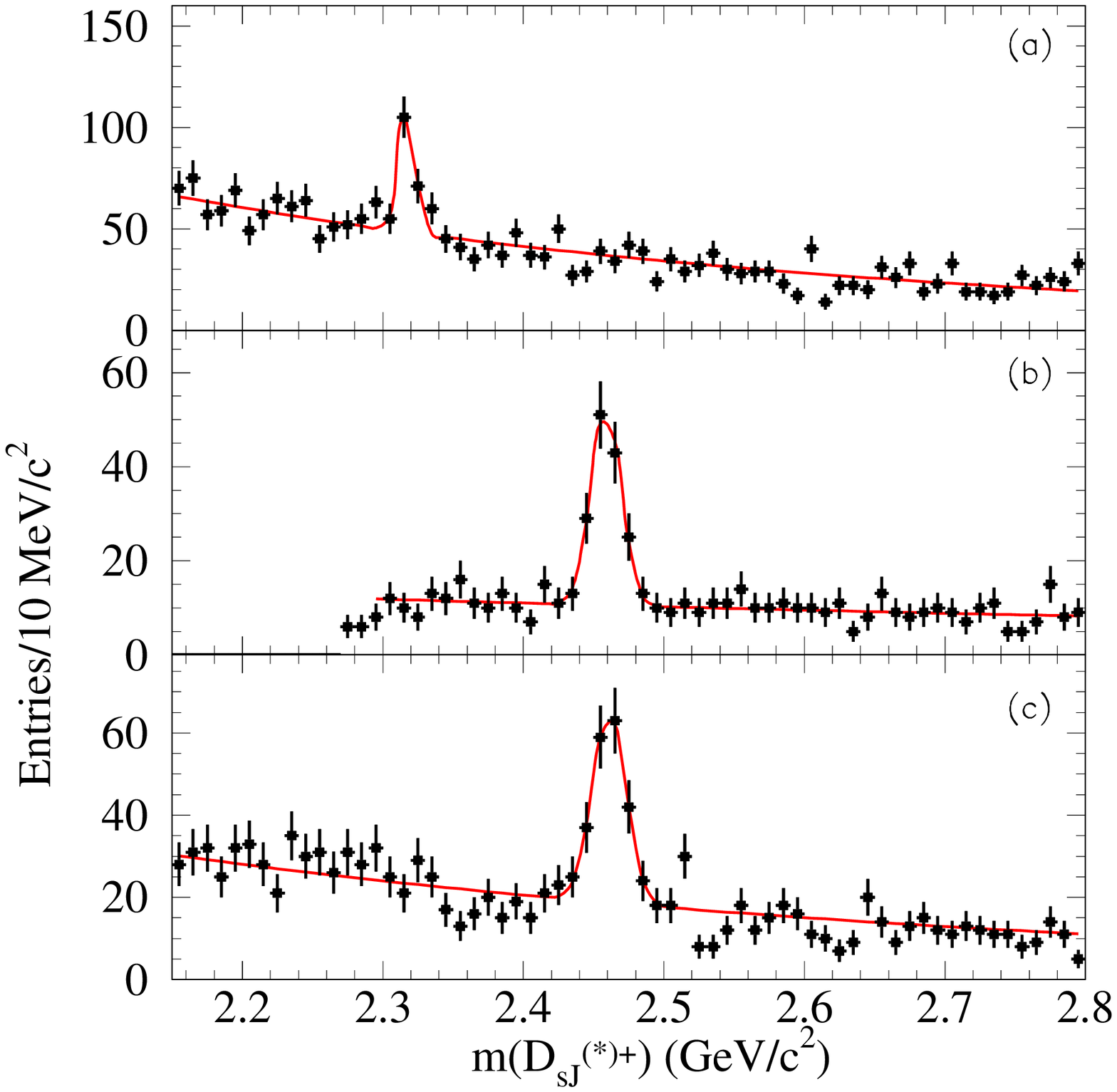}
 \end{center}
  \caption{
$\Delta E$ (left), \mes (center), and $m(\dsjp)$ (right) spectra
for the $B\to \dsjp \Db^{(*)}$ candidates: (a) $\dssz\rightarrow
D^{+}_{s} \piz$, (b) $\dsso\rightarrow D^{*+}_{s} \piz$, and (c)
$\dsso\rightarrow D^{+}_{s}\gamma$.  Data points in each plot show 
the distribution of one variable in the signal regions of the other two, 
as defined in the text.
For \DeltaE and \mes, the cross-hatched histograms are from events 
from the $m(\dsjp)$ sideband regions defined in the text. For the $m(\dsjp)$ plots, only 
one $B$ signal candidate per event has been selected.  Curves correspond to fit results.
} \label{fig:sumfits}
\end{figure*}

The selected pairs of $D_s^{(*)+}$ and $\overline{D}^{(*)}$
candidates are combined with a photon  or a \piz to form $B$
candidates. The photon energy is required to be greater than
$100\mev$. The neutral pions are built from pairs of photons with
energies above $30\mev$ and an invariant mass between $115$ and
$150~\mevcc$. A mass-constrained kinematic fit is applied to all
the intermediate particles. In order to suppress combinatorial
background, we require the $\overline{D}^{(*)}\piz/\gamma$
invariant mass to be greater than 2.3\gevcc and 2.4\gevcc for \Db
and \Dstarb final states, respectively. Events compatible with
being two-body decays $B\rightarrow D_s^{(*)+}\overline{D}^{(*)}$
are
rejected.

We define a $B$ signal region in terms of the beam energy
substituted mass, $\mes \equiv \sqrt{s/4-p^{*\ 2}_B}$, and the
difference between the reconstructed energy of the $B$ candidate
and the beam energy, $\DeltaE \equiv E^*_B-\sqrt{s}/2$, where
$\sqrt{s}$ is the total energy in the \FourS center-of-mass frame
and $E^*_B$ ($p^*_B$) is the energy (momentum) of the $B$
candidate in the same frame. We require $5.272<\mes<5.288 \gevcc$
and $|\DeltaE|<32(40)\mev$ for \piz ($\gamma$) final states.
The width of the signal box is approximately $\pm 3\sigma$ in
\mes and $\pm 2\sigma$ in \DeltaE. We also define in the
$D_s^{(*)+} \piz/\gamma$ mass spectra a signal region
$|m(D_s^{(*)+}\piz/\gamma)-m(\dsjp)|<2.5\sigma$
 and a sideband region from $4\sigma$ to $ 12\sigma$ away from 
the nominal value, with $m(\dsjp)=2.317\gevcc$ ($2.460\gevcc$) for 
\Ds\piz (\Dss\piz, \Ds$\gamma$). The resolution $\sigma=8\mevcc$ 
($12\mevcc$) for \piz ($\gamma$) final states is obtained from simulated signal events.

The \DeltaE, \mes, and $D^{(*)+}_s\piz$ or $D_s^+\gamma$ mass
spectra of the selected events are shown in Fig.~\ref{fig:sumfits}
for each of the three $\dsjp$ final states after combining the
charged and neutral $B\rightarrow \dsjp \Db^{(*)}$ modes and
summing over all the $\Db^{(*)}$ and $D_s^{(*)+}$ decays.
 Data points in each plot show the distribution of one variable in 
the signal regions of the other two. We also show (cross-hatched
 histograms) the \DeltaE and \mes spectra of events in the \dsjp\
 sidebands.

\begin{table}[htb]
  \centering
  \caption{Event yields, reconstructed {$\dsjp$} masses  and
  resolutions in each final state for $B\to\dsjp\Db ^{(*)}$ decays.
  }
  \label{tab:sumfits}
\begin{tabular}{llccc} \hline\hline
  \multicolumn{2}{c}{Decay mode}&  Yield & $m(\dsjp)$ & $\sigma_{m}$ \\
  \multicolumn{2}{c} {} & & [\mevcc]&  [\mevcc]\\ \hline
  {$\dssz\overline{D}^{(*)}$} &{$[D_{s}^{+} \piz]$}
    & $88\pm 17$& $2317.2\pm 1.3$ &  $5.9 \pm 1.4$
  \\
  {$ \dsso\overline{D}^{(*)}$} &{$[D_{s}^{*+} \piz]$}
  & $112\pm 14$ & $2458.9\pm 1.5$ & $10.8\pm 1.3$ \\
  {$\dsso\overline{D}^{(*)}$} &{$[D_{s}^{+} \gamma\;]$}
   & $139\pm 17$& $2461.1 \pm 1.6$ & $12.1 \pm 1.6$
  \\ \hline\hline
\end{tabular}
\end{table}

 Only one $B$ signal candidate per event,  based on the
smallest $|\DeltaE|$, is entered in the $D^{(*)+}_{s}\piz$ and
$D_{s}^{+}\gamma$ mass spectra and kept for further analysis. 
The $\dsjp$ yields, masses, and
resolutions, obtained from fitting a Gaussian signal function
and an exponential background to these 
 spectra, are given in Table~\ref{tab:sumfits}. The measured resolutions are
compatible
 with expectations from the simulation, assuming zero
 intrinsic width for $\dsjp$. We have also confirmed that the yields obtained
 from fits to the \mes and \DeltaE spectra  are in good agreement with
 the yields fitted from the \dsjp\ mass spectra.

\begin{figure*}[htb!]
\begin{center}
 \hspace*{-0mm}\includegraphics[width=6.2 cm,height=8cm]{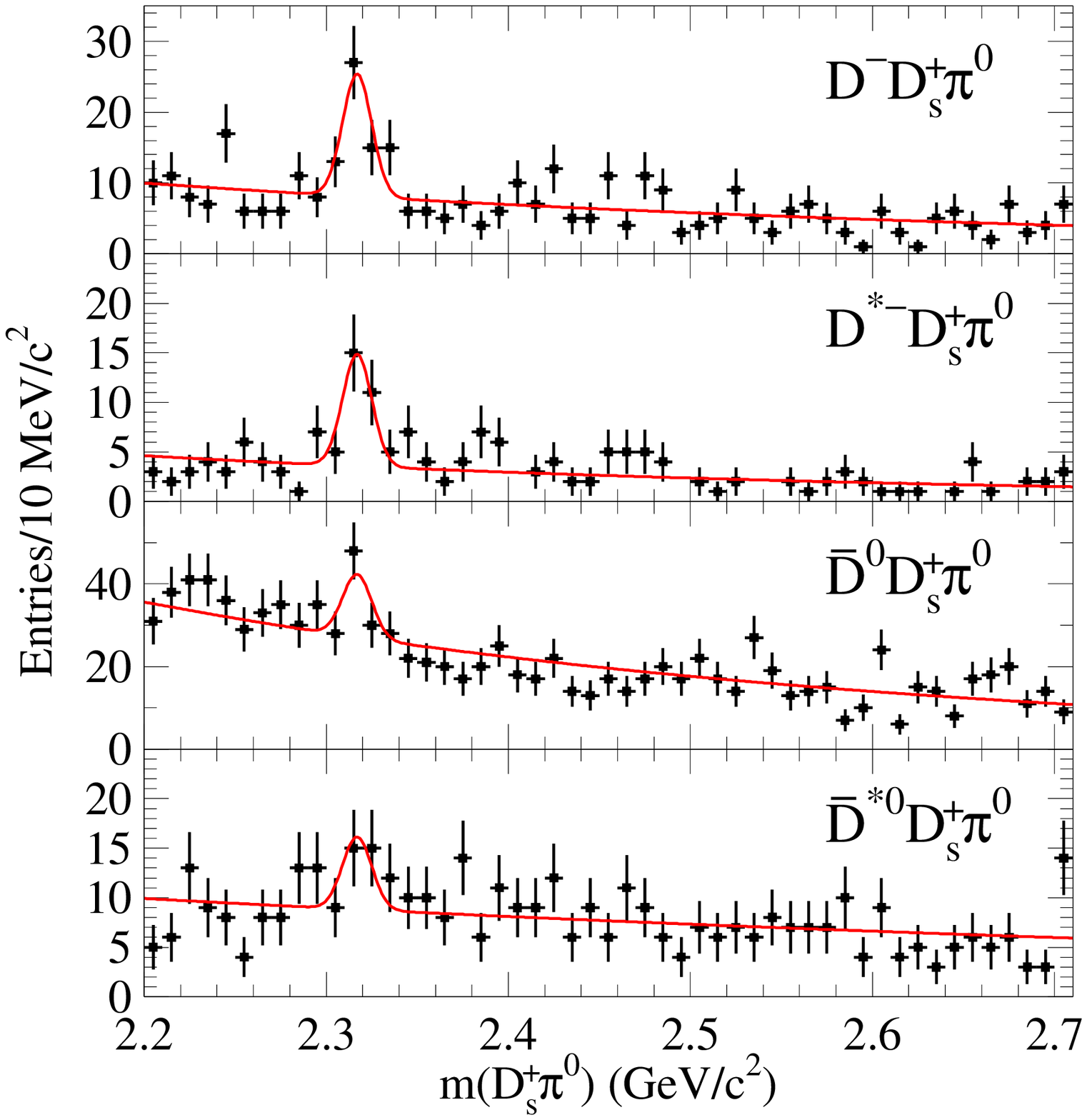}
 \hspace*{-5mm}\includegraphics[width=6.2 cm,height=8cm]{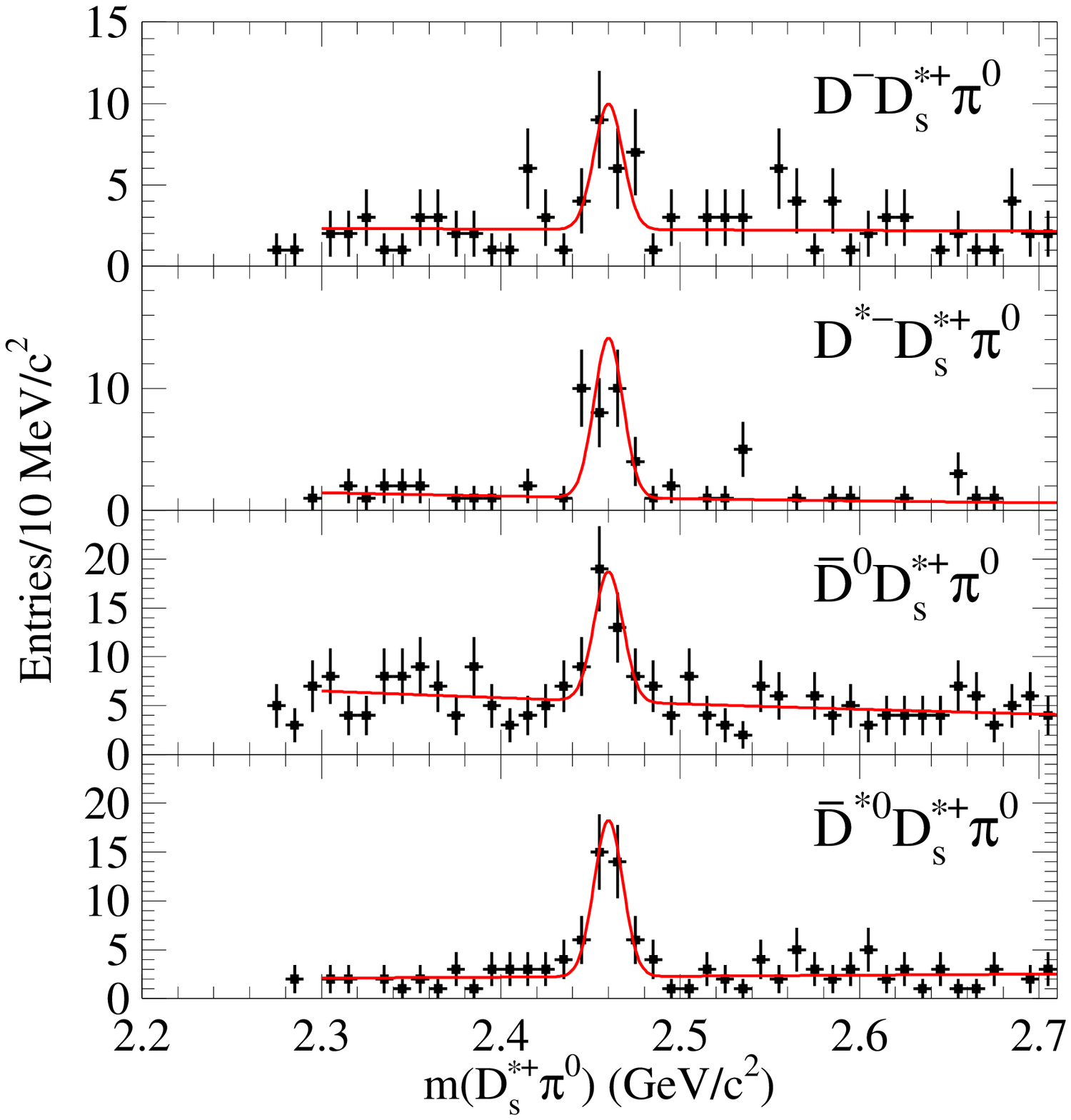}
 \hspace*{-5mm}\includegraphics[width=6.2 cm,height=8cm]{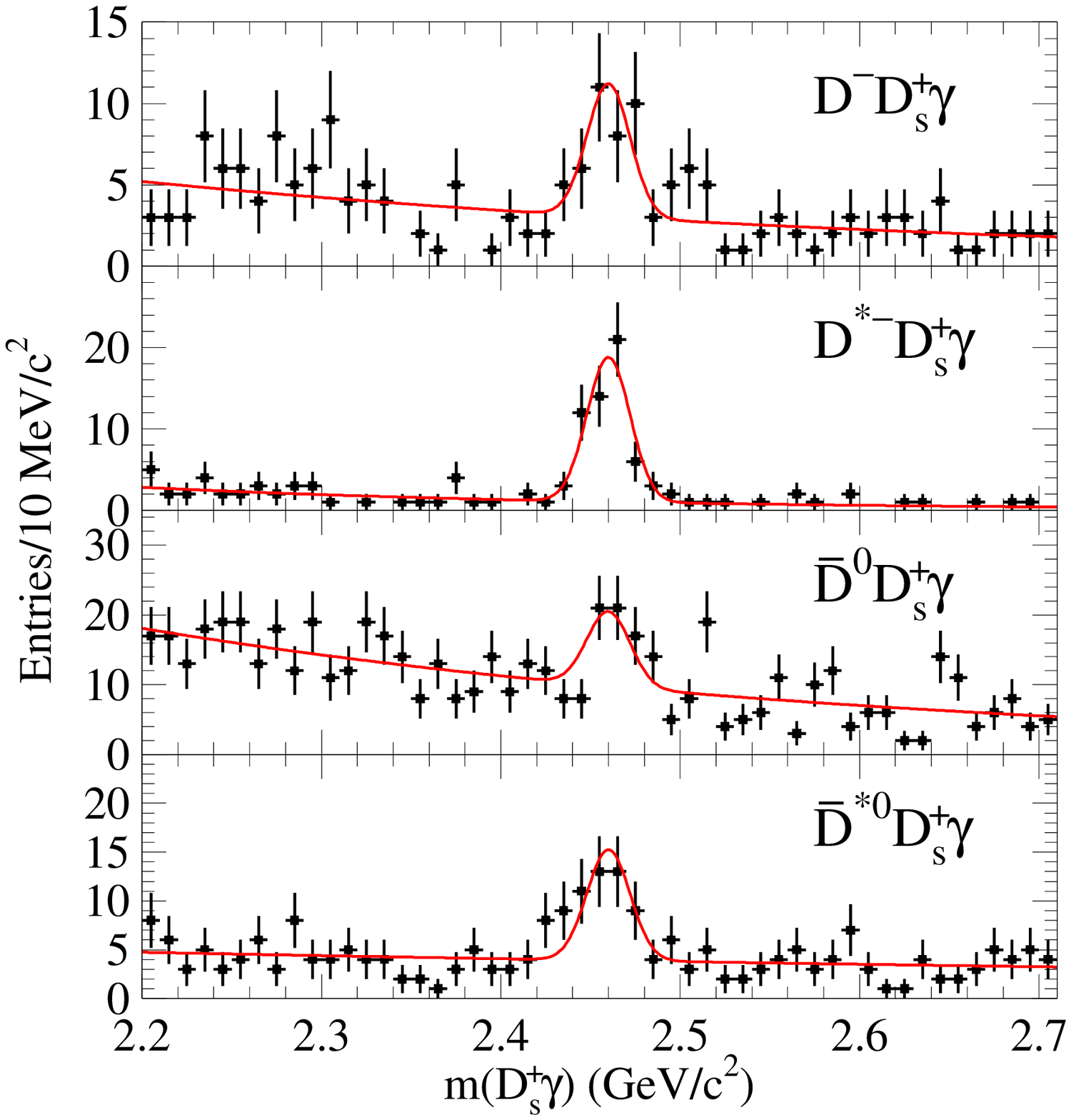}
 \end{center}
\caption{ $D_s^+\piz$ (left), $D_s^{*+}\piz$ (center), and
$D_s^+\gamma$ (right) mass spectra of the selected $B$ signal candidates
for the 12 $\Db^{(*)} \dsjp$ final states. Curves are the results
of the fits described in the text.}\label{fig:results}
\end{figure*}

The branching fraction measurement is based on the individual
{$D_s^+\piz$, $D_s^{*+}\piz$}, and {$D_s^+\gamma$} mass spectra for
each of the twelve {$D_{s}^{(*)+} \overline{D}^{(*)}\piz/\gamma$}
final states. As shown in Fig.~\ref{fig:results}, signals for
$B\rightarrow \dsjp \Db^{(*)}$ are observed in all channels. The
results of likelihood fits to these distributions, using a
Gaussian signal and an exponential background function, are
overlaid. In these fits, the Gaussian mean value is fixed to 2317 
(2460)\mevcc for $\dssz$ ($\dsso$). Its width is fixed to 8 (12)
\mevcc for \piz ($\gamma$) final states, as obtained from the
simulation and confirmed with the data (Table~\ref{fig:sumfits}). The
\dsjp\ event yields and the statistical significances  are listed
in Table~\ref{tab:br1st}. The significance is defined as
$\sqrt{-2\ln({\cal L}_0/{\cal L}_{max})}$, where ${\cal L}_{max}$
and ${\cal L}_{0}$ are the likelihood values with the nominal and
with zero signal yield, respectively. A significance larger than 4
is observed for 10 of the 12 modes.

From the $\dsjp$ event yields in the data, we compute cross-feed-corrected 
branching fractions,  using the signal efficiency and
the relative contributions from cross-feed between the different
$\dsjp$ decay modes as obtained from simulated signal events. The
resulting branching fractions are given in Table~\ref{tab:br1st},
together with the efficiencies, including the intermediate
branching fractions, and the internal cross-feed contributions.

\begin{table*}[htb]
  \centering
  \caption{Event yields (including internal cross-feed contributions), 
number of events attributed to internal cross-feed, efficiencies (including intermediate branching fractions), and final branching 
fractions, ${\mathcal B}$, for $B\to\dsjp\Db ^{(*)}$ decays. The first error on ${\cal B}$ 
is statistical, the second is systematic, and the third is from the \Db and \Ds branching 
fractions.} \label{tab:br1st}
\begin{tabular}{l@{$\rightarrow$}llr@{$\,\pm\,$}r@{.}lcccr@{.}l}
  \hline\hline
  \multicolumn{3}{c}{$B$ mode}
  & \multicolumn{3}{c}{Yield}  & Cross-feed & Efficiency $(10^{-4})$& ${\cal B}(10^{-3})$ & \multicolumn{2}{c}{Significance} \\
  \hline
\rule[0mm]{0mm}{1.1em}$B^0$&$ D_{sJ}^{*}(2317)^+D^{-}$ &[$D_{s}^{+} \piz$]
& $34.7$&8&0 & 0.3 & 1.6 & $1.8 \pm 0.4\pm 0.3 ^{+0.6}_{-0.4}$ &\hspace*{5mm}  5&5 \\
$B^0$&$ D_{sJ}^{*}(2317)^+D^{*-}$ &[$D_{s}^{+} \piz$]
& $23.5$&6&1 & 0.0 & 1.3 & $1.5 \pm 0.4\pm 0.2 ^{+0.5}_{-0.3}$ &  5&2 \\
$B^+$&$ D_{sJ}^{*}(2317)^+\overline{D}^{0}$ &[$D_{s}^{+}
\piz$]
& $32.7$&10&8 & 0.3 & 2.6 & $1.0 \pm 0.3\pm 0.1 ^{+0.4}_{-0.2}$ &  3&1 \\
$B^+$&$ D_{sJ}^{*}(2317)^+\overline{D}^{*0}$ &[$D_{s}^{+}
\piz$]
& $17.6$&6&8 & 7.2 & 1.0 & $0.9 \pm 0.6\pm 0.2 ^{+0.3}_{-0.2}$ &  2&5 \\
$B^0$&$ D_{sJ}(2460)^+D^{-}$ &[$D_{s}^{*+} \piz$]
& $17.4$&5&1 & 0.1 & 0.5 & $2.8 \pm 0.8\pm 0.5 ^{+1.0}_{-0.6}$ &  4&2 \\
$B^0$&$ D_{sJ}(2460)^+D^{*-}$ &[$D_{s}^{*+} \piz$]
& $26.5$&5&7 & 0.0 & 0.4 & $5.5 \pm 1.2\pm 1.0 ^{+1.9}_{-1.2}$ &  7&4 \\
$B^+$&$ D_{sJ}(2460)^+\overline{D}^{0}$ &[$D_{s}^{*+}
\piz$]
& $29.0$&6&8 & 2.2 & 0.8 & $2.7 \pm 0.7\pm 0.5 ^{+0.9}_{-0.6}$ &  5&1 \\
$B^+$&$ D_{sJ}(2460)^+\overline{D}^{*0}$ &[$D_{s}^{*+}
\piz$]
& $30.5$&6&4 & 2.5 & 0.3 & $7.6 \pm 1.7\pm 1.8 ^{+2.6}_{-1.6}$ &  7&7 \\
$B^0$&$ D_{sJ}(2460)^+D^{-}$ &[$D_{s}^{+} \gamma$]
& $24.8$&6&5 & 0.5 & 2.6 & $0.8 \pm 0.2\pm 0.1 ^{+0.3}_{-0.2}$ &  5&0 \\
$B^0$&$ D_{sJ}(2460)^+D^{*-}$ &[$D_{s}^{+} \gamma$]
& $53.0$&7&8 & 0.1 & 1.9 & $2.3 \pm 0.3\pm 0.3 ^{+0.8}_{-0.5}$ & 11&7 \\
$B^+$&$ D_{sJ}(2460)^+\overline{D}^{0}$ &[$D_{s}^{+}
\gamma$]
& $31.9$&9&0 & 1.4 & 4.1 & $0.6 \pm 0.2\pm 0.1 ^{+0.2}_{-0.1}$ &  4&3 \\
\rule[-0.5em]{0mm}{1.5em}$B^+$&$ D_{sJ}(2460)^+\overline{D}^{*0}$ &[$D_{s}^{+}
\gamma$]
& $34.6$&7&6 & 6.5 & 1.7 & $1.4 \pm 0.4\pm 0.3 ^{+0.5}_{-0.3}$ &  6&0 \\
  \hline\hline
\end{tabular}
\end{table*}

The dominant systematic errors come from the tracking efficiency 
(1.3\% per track), $\gamma$ and \piz efficiencies (2.5\% per 
$\gamma$), the $\Delta m$ requirement on the \Dss and \Dstarz selections
($\approx 5\%$ per $D^*$), efficiency of the \DeltaE requirement
($\approx 6\%$), \dsjp\ mass resolutions assumed in the fits (5 to
10\%), and background fitting model (5\%).
We assume equal production rates for $\Bp\Bm$ and $\Bz\Bzb$ pairs
and do not include a systematic error related to this assumption.
The errors from the individual $\Db^{(*)}$ and $D_s^{(*)}$
branching fractions, as taken from \cite{ref:pdg2004}, are given
separately (Table~\ref{tab:br1st}). They are dominated by the 25\%
relative error on ${\cal B}(D_s^+\to \phi \pip)$.

From the measured branching fractions for {$B \to D_{sJ}^+(2460)\overline{D}^{(*)}$} in the $D_s^{*+}\piz$ and in the
$D_s^{+}\gamma$ final states, we compute the ratio
$$\frac{{\cal B}(D_{sJ}(2460)^+ \to D_s^{+}\gamma)}{{\cal B}(D_{sJ}(2460)^+ \to
D_s^{*+}\piz)} = 0.274\pm 0.045 \pm 0.020,$$
where the first and second uncertainties are statistical and systematic respectively. 
This is compatible with the prediction from~\cite{Bardeen:2003kt}.

\begin{figure}[htb!]
 \begin{center}
  \hspace*{-5mm}\includegraphics[width=6.0cm]{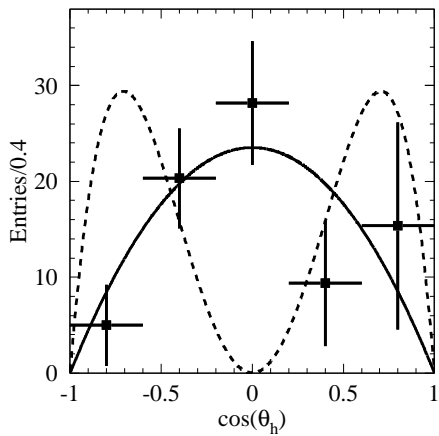}
 \end{center}

  \caption{Helicity distribution obtained from $m(D_s\gamma)$ fits in bins of $\cos(\theta_h)$
           for data (points) in comparison with the expectations for a $\dsso$ spin
           $J=1$ (solid line) and $J=2$ (dashed line), respectively, after normalizing the predicted spectra
           to the data.
} \label{fig:helicity}
\end{figure}

We perform a helicity analysis of the $\dsso$ state, using the
decays $B^+\rightarrow \dsso \overline{D}^0$ and
$B^0\rightarrow \dsso D^- $, with $\dsso \rightarrow D^+_s\gamma$.
 The helicity angle $\theta_h$ is defined as the angle between the 
$\dsjp$ momentum in the $B$-meson rest frame and the $D_s$ momentum 
in the $\dsjp$ rest frame. Since the $\overline{D}\gamma$ mass is
correlated with the helicity angle, the selection requirement
$m(\overline{D}\gamma)>2.3\gevcc$  is omitted for the angular
analysis. We perform $m(D_s\gamma)$  fits for five different
$\cos(\theta_h)$ regions, using the same fit functions and 
parameter values 
as in the corresponding branching fraction measurements. 

The resulting angular distribution,
after applying corrections for detector acceptance and selection
efficiency, is shown in Fig.~\ref{fig:helicity}. The  predicted
spectra for two different assumptions for the $\dsso$ spin, which
have been normalized to the data, are overlaid. We exclude the
$J=2$ hypothesis ($\chi^2$/n.d.f.=36.4/4) and find good agreement
with $J=1$ ($\chi^2$/n.d.f.=4.0/4). A \dsso\ spin $J=0$ is ruled
out by parity and angular momentum conservation in the decay
$\dsso \rightarrow D^+_s\gamma$.

In summary, we have observed and measured the branching fractions
for the decays $B \to D_{sJ}^{*}(2317)^+\Db^{(*)}$
($D_{sJ}^{*}(2317)^+\to \Ds \piz$) and $B \to 
D_{sJ}(2460)^+\Db^{(*)}$ ($D_{sJ}(2460)^+\to \Dss \piz$, $\Ds
\gamma$). The modes involving a \Dstarb have been seen for the
first time. The angular analysis of the decay $B \to
D_{sJ}(2460)^+\Db$ with $D_{sJ}(2460)^+\to \Ds \gamma$ excludes
$J^P=2^+$ and supports the hypothesis that the $D_{sJ}(2460)^+$ is
a $J^P=1^+$ state.

We are grateful for the excellent luminosity and machine conditions
provided by our \pep2\ colleagues, 
and for the substantial dedicated effort from
the computing organizations that support \babar.
The collaborating institutions wish to thank 
SLAC for its support and kind hospitality. 
This work is supported by
DOE
and NSF (USA),
NSERC (Canada),
IHEP (China),
CEA and
CNRS-IN2P3
(France),
BMBF and DFG
(Germany),
INFN (Italy),
FOM (The Netherlands),
NFR (Norway),
MIST (Russia), and
PPARC (United Kingdom). 
Individuals have received support from CONACyT (Mexico), A.~P.~Sloan Foundation, 
Research Corporation,
and Alexander von Humboldt Foundation.

\end{document}